# Physics and chemistry of the Purcell's alignment

A. Lazarian
*DAMTP, Silver Street, University of Cambridge, U.K.*★

**ABSTRACT**
Paramagnetic alignment of suprathermally rotating grains is discussed in view of recent progress in understanding subtle processes taking place over grain surface. It is shown that in typical ISM conditions, grains with surfaces of amorphous $H_2O$ ice, defected silicate or polymeric carbonaceous material are likely to exhibit enhanced alignment, while those of aromatic carbonaceous material or graphite are not. The critical grain sizes and temperature for the Purcell's alignment are obtained and preferential alignment of large grains is established.

## 1 INTRODUCTION

The paramagnetic mechanism of grain alignment (Davis & Greenstein 1951) is believed to be too weak to account for observations unless enhanced imaginary part of grain magnetic susceptibility is assumed (Jones & Spitzer 1967, Mathis 1986). Physically this weakness means that thermal disorientation due to chaotic atomic bombardment overwhelms orientation effects of dielectric grains with normal paramagnetic susceptibility rotating in the Galactic magnetic field. To diminish the effect of thermal disorientation it was suggested in Purcell (1975, 1979) that grains rotate suprathermally, i.e. with energy orders of magnitude greater than any temperature in the system times $k$ (the Boltzmann constant).

It may be shown that among various causes of suprathermal rotation those associated with corpuscular and radiative fluxes result in the Gold-type alignment (Lazarian 1994a, 1994b); its time-scale is orders of magnitude less than that of paramagnetic relaxation. Therefore, 'chemically driven' suprathermal rotation is required by the Purcell's mechanism. The estimates in Spitzer & McGlynn (1979) showed that if high rates of grain resurfacing are considered, it still improves the alignment only marginally. Thus, straightforward attempts to explain observations through the Purcell's mechanism (Aanestad & Greenberg 1983) had rather limited success (Spitzer 1993).

The major difficulties of the paramagnetic alignment can be summarised as follows.
(i) The paramagnetic alignment is too weak if ordinary paramagnetism is involved and 'standard' values of magnetic field and gaseous density are used (Spitzer & McGlynn 1979, Mathis 1986, Chlewicki & Greenberg 1990, Greenberg 1993, Spitzer 1993).
(ii) Observations indicate that grains of different chemical composition exhibit different degree of alignment (Mathis 1986, Whittet 1992) while this does not follow from the theory (see Jones & Spitzer 1967).
(iii) Large grains produce more polarization than small ones, while the Davis-Greenstein mechanism tends to go the other way (Mathis 1986, Kim & Martin 1994, Mathis 1994).

In our recent paper on the alignment of fractal grains (Lazarian 1994c), we were mostly concerned with the first difficulty mentioned above. This papers studies details of grain surface physics and chemistry to address the other two problems. Such a study would not be possible until quite recently as it is based on new results obtained in the field (Duley 1993, Duley & Williams 1993).

The structure of the paper is as follows. In section 2 we discuss the physics and chemistry of suprathermal rotation caused by $H_2$ formation over grain surfaces, whereas the paramagnetic relaxation is discussed in section 3. The ways of reconciling theory and observations are discussed in section 4 and two tests for the theory are suggested in section 5.

## 2 'CHEMICALLY DRIVEN' SUPRATHERMAL ROTATION

Although the interest to grain chemistry in view of grain alignment can be tracked back to Purcell & Spitzer's paper of 1971, it was only in Purcell (1975, 1979) that the concept of suprathermal rotation of grains driven by $H_2$ formation was introduced. These pioneering papers were based on the level of knowledge about grain chemistry achieved at that time. New discoveries in this rapidly growing area make us to update some of the ideas put forward by Purcell. This

★ Present address: Dept. of Astronomy, University of Texas, Austin, TX 78712, U.S.A.

update does not alter the core of the Purcell's concept, but shows that the 'chemically driven' suprathermal rotation is not universally applicable, e.g. to grains of different chemical composition.[†]

It is shown in Duley & Williams (1993) that $H_2$ molecules are formed over aromatic carbonaceous surfaces in low-energy state as the PAH.2H complex shares all the energy of stabilisation of $H_2$ (4.5 eV) among its degrees of freedom'. This prediction entails that such grains should not rotate suprathermally due to $H_2$ formation.

The situation is different for H atoms weakly adsorbed by water ice or polymeric hydrocarbon. The binding is weak, its energy is not expected to exceed 0.1 eV (Lee 1972, Leitch-Devlín & Williams 1984, Zhang & Buch 1990, Buch & Zhang 1991). $H_2$ molecules formed on such sites stabilise into very high vibrational states (Duley & Williams 1993). A portion of vibrational energy ($\sim 0.1 - 0.2$ eV) is likely to be transferred to the center-of-mass translational motion (Tielens & Allamandola 1987) at approximately $10^7$ s$^{-1}$ rate (Lucas & Ewing 1981). Thus an injection of high speed $H_2$ molecules is expected from icy or polymeric hydrocarbon grain surfaces. Similarly using results in Duley & Williams (1986) it is possible to show that $H_2$ molecules formed over defected silicate surfaces are also expected to have high kinetic energies. For the hydrogenate amorphous carbon (HAC) model of interstellar dust (Jones et al. 1990) which consists of a mixture of polymeric, aromatic and diamond-like regions of carbon separated by voids, the situation is more involved. $H_2$ formation is expected to be efficient in the sense of recoils deposited with grains for polymeric and probably diamond-like regions while aromatic regions are expected to eject $H_2$ with low kinetic energy.

The ejection of $H_2$ molecules with high kinetic energy $E$ results in torques applied to a grain. Everywhere below we assume $E \approx 0.2$ eV. This means that our approach is applicable to suprathermal rotation of grains with only icy, polymeric carbonaceous or silicate surfaces.

It was shown in Purcell (1979) that internal dissipation of energy within grains, mainly due to the Barnett relaxation,[‡] suppresses rotation around any axis but the axis of the greatest inertia on the time-scale $\sim 10^7/\eta$ s, where $\eta$ is the ratio of grain rotational energy to the equipartition energy $\sim kT$. Thus grains rotate around their major axes of inertia that we denote here the $z$-axis. Therefore the problem becomes one dimensional as the two other components of the torque contribute only to insignificant nutations. The number of $H_2$ molecules ejected per second from an individual site is $\sim \gamma_1 l^2 v_H n_H \nu^{-1}$, where $\gamma_1$ the portion of H atoms with velocity $v_H$ and concentration $n_H$ adsorbed by the grain, while $\nu$ is the number of active sites over the grain surface and $l$ is the characteristic grain size which coincides with the diameter for a spherical grain. Then the mean square of a residual torque from an $H_2$ molecule of mass $m_{H2}$ is

$$\langle M_z^2 \rangle \approx \frac{\gamma_1^2}{32} l^6 n_H^2 m_{H2} v_H^2 E \nu^{-1} \qquad (1)$$

which spins up grains to angular velocities limited only by friction forces, i.e.

---

[†] We will consider rotation induced by $H_2$ formation, although some other chemical reactions might also contribute to the grain rotation in dark clouds, where the abundance of atomic hydrogen is low.

[‡] A quantitative treatment through solving Fokker-Planck equations is given in Lazarian (1994a).

$$\Omega = \langle M_z^2 \rangle^{1/2} \frac{t_d}{I_z} \qquad (2)$$

where $I_z$ is the $z$ component of the momentum of inertia, and $t_d$ is the rotational dumping time (Spitzer & McGlynn 1979):

$$t_d \approx 0.6 t_{gas} = 0.6 \frac{l \varrho_s}{n m v_1} \qquad (3)$$

where $t_{gas}$ is the time that takes a gain of density $\varrho_s$ to collide with gaseous atoms of the net mass equal to that of the grain, $n$ and $v_1$ denote the density and velocity of atoms, respectively, while $m$ is the mass of an individual atom. To obtain both numerical values and functional dependencies further on we will use quantities normalised by their standard values. To tell the normalised values we will denote them by corresponding letters with hats, e.g. $\hat{l}$ is equivalent to $l/(2 \cdot 10^{-5}$ cm) with $2 \cdot 10^{-5}$ cm as a standard value of a grain size. We consider 'standard' $\gamma_1 = 0.2$ and $v_{H2} = 5 \cdot 10^5$ cm s$^{-1}$. Speaking about diffuse clouds, we assume that all hydrogen there is in atomic form and therefore $n/n_H \equiv 1$ and $m/m_{H2} \equiv 0.5$. Note that sometimes the choice of 'standard' values is arbitrary, e.g. for the time being we assume the density of active sites $\alpha_{H2}$ to be $10^{11}$ cm$^{-2}$. However, we will see in section 3 that $\alpha_{H2}$ sensitively depends on subtle processes of poisoning and therefore 'standard' values may be misleading.

Using standard values of the parameters involved and assuming that the number of active sites is proportional to $l^2$, it is possible to write the following expression for the angular velocity

$$\Omega \approx 1.4 \cdot 10^8 \hat{\gamma}_1 \hat{\alpha}_{H2}^{-1/2} (\hat{l})^{-2} \text{ s} \qquad (4)$$

If grains are porous and fluffy as those found in the stratosphere (see fig. 7.3 in Williams 1993) the degree of suprathermal rotation can be limited provided that $H_2$ formation takes place within narrow irregular pores of grains and nascent molecules thermalize due to numerous collisions with pore walls. These processes are discussed in more detail in Lazarian (1994c), and here we mention this in view of implications that the suprathermal rotation can have on grain structure. The tensile stress within the spinning grain is of the order of $\varrho_s l^2 \Omega^2$ may reach $10^8$ dyn cm$^{-2}$. An interesting hypothesis by Wright (1994) is that suprathermal rotation limits the fractal dimension of grains may have the following consequences:
a) if grains have high fractal dimension, their degree of suprathermality is low and their fractal dimension does not change,
b) grains of aromatic carbonaceous material and graphite are not limited in their fractal dimension by suprathermal rotation,
c) grains with surfaces of defected silicate, amorphous $H_2O$ ice or polymeric carbonaceous material with a limited number of pores will have their fractal dimension decreased as a result of suprathermal rotation.

In short, silicate, ice and carbonaceous grains may have either large $d \to 3$ or low $d \to 2$ fractal dimension. One may also speculate that if $\alpha_{H2}$ is proportional to grain surface, $\Omega$ scales as $l^{-2}$ [see Eq. (4)], the tensile stress scales as $l^{-2}$ as well and therefore small grains are more vulnerable to centrifugal disruption, which may make them more spherical as compared to larger grains. Adopting a model of a grain as consisting of small particles jumbled together (Mathis 1990), one may say that these conglomerates are more subject to disruption if they are small. However, this problem needs further investigation and we will not proceed with its discussion here.

Grain dynamics depends not only on the number of active sites but also on their distribution over grain surface. Both accretion of a new monomolyer and poisoning of active sites cause occasional 'crossovers' (Spitzer & McGlynn 1979) which considerably effect grain alignment. If these crossovers occur on the time-scale $t_L$, one should multiply our estimates of $\Omega$ given by Eq. (2) by $t_L^{1/2}/(t_L + t_d)^{1/2}$.

## 3 THE PURCELL'S MECHANISM

### 3.1 Paramagnetic relaxation

A paramagnetic grain rotating in the external magnetic field **B** is subjected to dissipative torques (Davis & Greenstein 1951, see also Jones & Spitzer 1967, Purcell & Spitzer 1971, Roberge et al. 1993). In the grain coordinate system, the component of **B** perpendicular to the rotation axis appears as a rotating field. The magnetic moment produced by the imaginary part of magnetic susceptibility $\chi''$ causes rotation of the vector of angular momentum towards the direction of **B**. For low frequencies, $\chi''$ of paramagnetic materials is proportional to the component of the angular velocity perpendicular to **B** up to frequencies $\omega_c \approx 10^7 - 10^8$ s$^{-1}$. If $\Omega \gg \omega_c$, $\chi''$ becomes independent of $\Omega$ and approaches the static susceptibility. In short, large grain angular velocities increase the characteristic time of paramagnetic relaxation as $\chi''$ comes to its asymptotic values (compare Whittet 1992). The characteristic time of this relaxation is $t_r \approx 10^{14} q_1^{-1} \hat{l}^2 \hat{T}_s \hat{B}^{-2} K(\Omega) \hat{\varrho}$ s (Spitzer 1978), where grain temperature $T_s$ and density $\varrho$ are normalised by 10 K and 2 g cm$^{-3}$ respectively, while magnetic field is normalised by $2 \cdot 10^{-6}$ G, $q_1$ is the enhancement factor of the imaginary part of magnetic susceptibility suggested in Jones & Spitzer (1967), and $K(\Omega)$ is a function with the following asymptotics

$$K(\Omega) = \begin{cases} 1, & \Omega \ll \omega_c \\ \frac{\Omega}{\omega_c}, & \Omega \gg \omega_c \end{cases} \quad (5)$$

It was noted by Mathis (1994) that a low value of $B$ is used above as standard and there might be a substantial increase for $B^2$ term (see e.g. Boulares & Cox 1990, fig 3b, where $B^2/8\pi$ is shown in cgs units). 'However, many people would not be willing to go beyond about 3 micro G for the standard ISM field in the Galactic plane', he concludes. Later in section 4 we will study the possibility that enhanced values of magnetic field are *responsible for alignment*.

For simple estimates, one may consider that transition from one asymptotic to another is at $\Omega = \omega_c$. Within this simple model, it is possible to introduce a critical grain size $l_{cr\ m}$ when the imaginary part of paramagnetic susceptibility comes to saturation. For long-lived spin-up, i.e. $t_L \gg t_d$, Eqs. (4) and (5) provide

$$l_{cr\ m} = 10^{-5} \left( \frac{5.6 \cdot 10^8}{\omega_c} \right)^{1/2} \hat{\gamma}_1^{1/2} \hat{\alpha}_{H2}^{-1/4} \hat{v}_{H2}^{1/2} \text{ cm} \quad (6)$$

and for short-lived spin-up, i.e. $t_L \ll t_d$, it is $(t_L/t_d)^{1/4}$ as large.

For the original Davis-Greenstein process, $t_r$ should be compared with the time that takes a grain of mass $m$ to collide with gas atoms of the same total mass $t_{gas} = 2 \cdot 10^{12} \hat{l} \hat{\varrho} (\hat{v}_H \hat{n}_H)^{-1}$ s, where the gaseous density is normalised by 20 cm$^{-2}$ and the 'standard' thermal velocity of atoms $v_H$ is assumed to be $1.3 \cdot 10^5$ cm s$^{-1}$. Note, that in many cases the velocity $v$ is dominated by Alfvénic phenomena (see Aron & Max 1975, Elmegreen 1992, Myers 1987, 1994, Lazarian 1992, 1994a), the fact that makes $t_{gas}$ smaller.

While a grain rotates suprathermally, chaotic bombardment of atoms cannot alter its rotation. It was argued in Purcell (1979) that a quantity $t_x = 1.3(t_L + t_d)$ should be compared with $t_r$ to find the Rayleigh reduction factor $\langle \sigma \rangle$ introduced in Greenberg (1968), which relates the measures of polarization and grain alignment (see Whittet 1992). The expression of $\langle \sigma \rangle$ is given in Aanestad & Greenberg (1983) and here we just mention that $\langle \sigma \rangle$ depends on the following quantity (Spitzer & McGlynn 1979):

$$\delta_{eff} = \begin{cases} t_x (t_r)^{-1}, & F > 1 \\ t_x (F t_r)^{-1}, & F < 1 \end{cases} \quad (7)$$

where $F$ is the 'disorientation parameter' explicit analytical expression of which is obtained in the Appendix. One may establish a critical size $l_{cr\ f}$ after which $F$ becomes greater than unity [see Eq. (A11)]

$$l_{cr\ f} = 1.8 \cdot 10^{-5} \frac{(\hat{v}_{H2} \hat{\nu})^{1/3}}{\hat{\varrho}^{1/4} (\hat{n}_H \hat{v}_H \hat{\gamma}_1)^{1/12}} \quad (8)$$

where $\hat{\nu}$ is normalised by 100. If density of active sites is proportional to grain surface (we will see further on that this may not be the case):

$$l_{cr\ f} = 1.7 \cdot 10^{-5} \frac{\hat{\alpha}_{H2}^{1/4}}{\hat{\varrho}^{3/8} (\hat{n}_H \hat{\gamma}_1 \hat{v}_H)^{1/8}} \text{ cm} \quad (9)$$

For grains greater than this size, it is important to account for partial preservation of alignment during the crossovers.

In the limit of $t_L \ll t_{gas}$ and $l < l_{cr\ f}$, $\delta_{eff}$ coincides with the parameter of alignment for the Davis-Greenstein mechanism and is equal to $t_{gas}/t_r = 0.02 \ (\hat{l} \hat{T}_s \hat{v}_H \hat{n}_H)^{-1} \hat{B}^2 \ K(\Omega) q_1$.

Up to now we have discussed disorientation due to collisions with gaseous atoms. The disorientation due to radiation introduced in Purcell & Spitzer (1971) can be characterised by the time-scale $t_{rad} \approx 5 \cdot 10^{13} \hat{l} \hat{\varrho} \hat{T}_s^{-4}$ s.

### 3.2 Grain spin-up

To obtain sufficient alignment for $q_1 = 1$, it is required that $t_L \gg t_{gas}$. The difficulty on this way is the finite life time of active sites over grain surface.

A process that can limit the life time of active sites is resurfacing due to the accretion of a new monomolecular layer (Spitzer & McGlynn 1979). A grain of mass $m(l)$ grows as

$$\frac{\mathrm{d}m(l)}{\mathrm{d}t} = \frac{1}{4}S(l)vn_a m\xi_a \qquad (10)$$

where $\xi_a$ is the adsorption probability, a coefficient with the range of values from 0 to 1 for atoms of density $n_a$, velocity $v$ and mass $m$; $S(l) \sim \pi l^2$ is the geometric surface of the grain. Assuming that accreting atoms uniformly cover the entire physical surface of grains and substituting $\mathrm{d}m(l) = \varrho S_{ph}(l)\mathrm{d}l$, where $S_{ph}(l)$ is the physical surface area of grain, one obtains for the accreting time:

$$t_a = 4t_{gas}q_2 \left(\frac{\triangle h}{l}\right)\left(\frac{n_H}{n_a}\right)\left(\frac{m_H}{m}\right)\left(\frac{v_H}{v}\right)\frac{1}{\xi_a} \qquad (11)$$

where $\triangle h$ is the thickness of an adsorbed layer, and $q_2$ is $S_{ph}(l)/S(l)$ ratio. If $v_H$ and $v$ are due to thermal motions, then $\left(\frac{v_H}{v}\right) = \left(\frac{m}{m_H}\right)^{1/2}$ and $t_a$ becomes proportional to $\left(\frac{m_H}{m}\right)^{1/2}$. However, if non-thermal Alfvénic motions dominate, $v_H$ becomes equal to $v$ and $t_a$ is proportional to $\left(\frac{m_H}{m}\right)$.

We will see further on that oxygen is the most important agent poisoning the active sites. Other chemical elements either form compositions that desorb more easily than $H_2O$ or have a negligible abundance. As all heavy elements, oxygen is usually depleted in the gaseous phase. For instance, its depletion towards $\zeta$ Persei is -0.5, which gives $n_O/n_H \approx 10^{-3.7}$ (Cardelli et al. 1991). Assuming that a layer of $\triangle h \approx 3.7 \cdot 10^{-8}$ cm effectively covers active sites over grain surface (Aanestad & Greenberg 1983), one obtains $\triangle h/l \approx 1.85 \cdot 10^{-3}$. Therefore for the accretion time-scale for oxygen is $t_a \approx 8t_{gas}\xi_a^{-1}q_2$. The estimate for $t_a$ can be further increased if $\xi_a < 1$ and the desorption is present. The value of $\xi_a \sim 0.1$ was used in Spitzer (1978, p. 208) for illustrative purposes while the values 0.02 and 0.2 were reported for sticking probabilities of $C$ over silicate and graphite covered with a monolayer of $H_2O$ surfaces (Williams 1993). It is shown in Jones & Williams (1984) that the effective sticking probability varies as mantles begin to be deposited and as a consequence the time of establishing monolayer coverage may be increased by a factor of about three.

The products of grain surface chemistry are also subjected to desorption which increases the time of resurfacing. According to Jones & Williams (1984) the probability of retaining the products OH and $H_2O$ on the grains surface is about 0.7 in dark quiescent clouds, whereas NH may be predominantly released on formation (Wagenblast et al. 1993). Immediately after formation the reaction products are subjected to photodesorption. This process can be driven not only by UV quanta (Johnson 1982), but also by water ice 3 $\mu$m absorption (Williams et al. 1992). Note, that 3 $\mu$m radiation penetrates deep into dark clouds. The energy release associated with $H_2$ formation may be another cause of desorbing weakly bound molecules such as $CO$ in the vicinity of active sites (Duley & Williams 1993). This means that active sites can control to some extent resurfacing in their vicinity. In brief, we may expect to obtain $t_a$ of the order of $10\, t_{gas}q_2$ using rather conservative estimates for the above parameters.

However, such considerations are applicable only if heavy atoms form an inert monolayer rather than poisoning individual active sites. The thermal hopping of atoms has the characteristic time (Watson 1976)

$$t_h \approx \frac{1}{\nu_0}\exp\left(\frac{E_h}{kT_s}\right) \qquad (12)$$

where $\nu_0 \approx 10^{12}$ s$^{-1}$ is the characteristic frequency of vibration for an adsorbed particle and $E_h$ is the energy of potential barrier which is approximately one third of the total binding energy (although see Jaycock & Parfitt 1986). The total binding energy for oxygen may be assumed $\sim k \cdot 800$ K (Tielens & Allamandola 1988). In this case, the normalised $\hat{E}_h$ is equal to $E_h/(k \cdot 267$ K).

An adsorbed H atom scans the entire surface of a grain of $l = 2 \cdot 10^{-5}$ cm on the time-scale not exceeding $10^{-2}$ s. If it meets another H atom they are likely to form $H_2$ molecule on a time-scale much shorter than the time of adsorbing another hydrogen atom. Therefore if an oxygen atom is adsorbed after this $H_2$ formation took place, it is free to scan the grain surface unless a new H atom is adsorbed.

Consider an oxygen atom arriving at the grain surface. The sites of chemical adsorption can be either all filled with H atoms or be empty due to a recent $H_2$ formation. The number of empty sites is limited. The activation barrier of $10^3$ K and a width of $10^{-8}$ cm is usually assumed for reactions between physically and chemically adsorbed hydrogen atoms (Tielens & Allamandola 1988). This gives the time-scale for the reaction of the order $1.3 \cdot 10^{-6}$ s. The migration time-scale depends on the properties of the grain surface. Quantum tunneling corresponds to the time-scale of the order $10^{-12}$ s, although estimates of the order $3 \cdot 10^{-9}$ s can be also found in the literature (Tielens & Allamandola 1988). As we will see further, the above difference is crucial. Indeed, assuming the time for migration $10^{-12}$ s, one obtains that before reacting, a hydrogen atom has chances of visiting $\nu_{cr} = 10^6$ sites of chemical adsorption, while this estimate falls to just $\nu_{cr} \approx 500$ if the other value for the migration time-scale is assumed. In the former case a grain with high density of active sites will have not more than one empty active site of $H_2$ formation even if a grain is completely covered by active sites; in the latter case up to $\sim \nu/\nu_{cr}$ (if $\nu > \nu_{cr}$) active sites may be empty. These empty active sites are the prime targets of an O atom hopping over the surface. If O atom is trapped in an active site, it will stay there and will be eventually hydrogenated with the reaction products blocking the access to the site.

If we assume that the number of sites of physical adsorption is $N_{ph}$ and among $\nu$ active sites one is empty, the probability of an O atom not to fill it as a result of an individual hop is $1 - N_{ph}^{-1}$; this probability for $m_h$ hops decreases to $(1 - N_{ph}^{-1})^{m_h}$. If $m_h \gg 1$ and $N_{ph} \gg 1$, this expression can be approximated:

$$(1 - N_{ph}^{-1})^{m_h} \approx \exp\left(-\frac{m_h}{N_{ph}}\right). \qquad (13)$$

The number of hops $m_h$ before an O atom hydrogenated is proportional to the ratio of the time of adsorbing a hydrogen atom $t_c = 1.5 \cdot 10^3 (\hat{\gamma}_1 \hat{n}_H \hat{v}_H \hat{l}^2)^{-1}$ s to the time $t_h$ given by Eq. (12). If a time-scale for hydrogen diffusion is $\sim 10^{-12}$ s for a grain of $10^{-5}$ cm with high density of active sites hydrogen atom will scan the entire surface of the grain and will react with an oxygen atom. If the mobility of hydrogen is lower, the probability to hydrogenate O atom over a grain will be suppressed and $m_h$ will be increased by a factor $\exp(-\nu/\nu_{cr})$. Therefore for sufficiently high $\nu/\nu_{cr}$ ratio the majority of O atoms will eventually fill the empty active sites and the fact that the number of vacancies per grain increases from one to $\nu/\nu_{cr}$ only contributes to the effect. The exponential increase in poisoning of active sites when their number exceed $\nu_{cr}$ is likely to decrease their number to $\nu_{cr}$. The characteristic time of poisoning is as follows

$$t_p \approx 2\nu t_c \left(\frac{v_0}{v_H}\right)\left(\frac{n_H}{n_0}\right) \frac{1}{1-\exp(-m_h/N_{ph})}, \quad (14)$$

where $v_0$ is the velocity of O atoms with the number density $n_0$, the last factor is inversely proportional to the probability of O atom to fill an empty active site and the factor 2 reflects that in half of the cases O atoms will find that all sites are filled. For small $m_h/N_{ph}$ ratio, the exponent can be expanded into series and $t_p \approx 2\nu/m_h\ t_a$. In other words, for $t_p > t_a$, the number of active sites should not be less than $m_h$, which is for $\nu/\nu_{cr} < 1$ is equal to

$$m_h = \frac{t_c}{t_h} \approx 4 \cdot 10^3 (\hat{\gamma}_1 \hat{l} \hat{n}_H)^{-1} \hat{T}_g^{1/2} \exp\{-\hat{E}_h/\hat{T}_s\} \quad (15)$$

Therefore a grain of $2 \cdot 10^{-5}$ cm should have not less than $2 \cdot 10^3$ active sites for $t_p$ to be greater than $t_a$. This means that the time-scale for hydrogen migration should be less than $10^{-9}$ s. If $\nu = \nu_{cr} = 500$ as in example discussed above, $t_L$ becomes $0.25 t_a$ which corresponds to $\sim 2.5 t_{gas} q_2$. Although less than our estimate for the time of spin-up limited only by monomolecular layer accretion, this estimate provides a substantial increase of $t_L$ for irregular ($q_2 \gg 1$) grains. The functional dependence of $t_L$ as a function of grain fractal dimension obtained in Lazarian (1994c) also stays unaltered as $t_p \sim t_a$. However, the processes discussed here depend not only on the relative abundance of oxygen as it was assumed in Lazarian (1994c) but also on the grain temperature $T_s$ and the energy of the potential barrier $E_h$. Both values can be measured and therefore long-lived spin-up can be tested.

Consider, for instance, grains covered by $H_2O$ ice. Simulations in Buch & Zhang (1991) indicate that active sites (i.e. sites with binding energy in excess of $k \cdot 900$ K) cover around $10^{-2}$ of the entire grain surface that a 'standard' grain will have $\nu \sim 1.5 \cdot 10^4$ active sites. Data in table 2 in Tielens & Allamandola (1988) show that migration scales for hydrogen and oxygen are, respectively, $10^{-12}$ s and $10^{-2}$ s. For such high mobilities of hydrogen $\nu_{cr} > \nu$ and therefore $\nu$ should be used in Eq. (14). A substitution of these values into Eq. (15) provides $m_h \approx 1.5 \cdot 10^5$ and thus $t_L$ becomes ten times less than $t_a$. To obtain $t_L \gg t_d$, one needs assume either $q_2 \gg 1$ or $T_s < 10$ K. For $T_s = 15$ K, $m_h$ becomes $\sim 6 \cdot 10^7$ and long-lived spin-up becomes improbable. On the basis of this data it is evident that the Purcell's alignment of grains with surfaces covered with amorphous $H_2O$ ice is limited to a rather narrow range of grain temperatures if our data for $\nu$ and $E_h$ are correct.

The critical size of the grain corresponds to $t_c$ becoming equal to the time for an oxygen atom to scan the grain surface. The time $t_c$ varies as $\gamma^{-1} l^{-2}$, the time for O atom to migrate over the grain surface as $q_2 l^2$; hence the critical size is

$$l_{cr\ p} \approx 6 \cdot 10^{-6} (q_2 \hat{n}_H \hat{\gamma}_1)^{-\frac{1}{4}} \hat{T}_g^{\frac{1}{8}} \exp\left\{-0.25 \hat{E}_h/\hat{T}_s\right\} \text{ cm} \quad (16)$$

If $T_s$ increases or $E_h$ decreases, rapid poisoning becomes essential for grains of all sizes. At the first glance, $l_{cr\ p}$ seems to be insensitive to the atomic hydrogen concentration: even a steady-state value of $n_H \sim 1 \text{cm}^{-3}$ (see e.g. Duley & Williams 1984) the critical size becomes only twice as large as the one given by Eq. (16). However, this is only true if the atomic hydrogen concentration is not less than that of oxygen. Otherwise, there is too little of hydrogen to immobilise poisoning O atoms.

For grains of size less than $l_{cr\ p}$, the crossover time [see Eq. (14)] is in order of magnitude $t_L \approx \nu t_c (n_H/n_0)$. This time is $\sim 10^8$ for standard values of parameters adopted in the paper and is much less than the corresponding $t_{gas} \sim 5 \cdot 10^{11}$ s. If the number of active sites is proportional to $l^2$, than $t_L/t_{gas} \sim 10^{-3}$ will not change as the grain becomes smaller. Consequently, $t_x$ becomes equal to $t_{gas}$ and the improvement of alignment as compared to Davis-Greenstein process is marginal.

## 4 PROBLEMS AND SOLUTIONS

### 4.1 Degree of alignment

The most efficient polarising medium conceivable would contain perfectly aligned infinite cylinders with their axes perpendicular to the line of sight. For such a model Mie calculations show that the ratio of polarization to extinction is of the order 0.3. At the same time, the observational upper limit for the same ratio is $\leq 0.064$ (see Serkowski et al. 1975). The difference between the two limits stems from the fact that realistic grains are irregular, moderately elongated particles and it is only a fraction of grains that is aligned. Nevertheless, it is difficult to account for the observations if $\delta_{eff}$ does not exceed unity.

Our estimates of $\delta_{eff}$ give $\sim 0.3$ for standard $l$ and $t_L = 10 t_{gas}$ (see section 2). Although 10 times greater than the initial value it is not sufficient to explain the polarization. One of the ways to increase $t_L$ is by taking into account photodesorption (see Johnson 1982). Indeed, as the characteristic time of photodesorption is around $10^9$ s for unshielded regions of diffuse clouds, which is usually much less than the time of resurfacing, this may considerably increase $t_L$. However, the correlation of polarization with the strength of 3.1 $\mu$m water ice feature (see Joyce & Simon 1982) testify that the photodesorption cannot provide a universal solution (see also Lee & Draine 1985).

One may question 'standard' values for magnetic field that is responsible for grain alignment, while leaving the 'standard' value of the mean magnetic field intact. Indeed, we observe the cumulative polarization and extinction along the line of sight, which presumably crosses quite different regions. If aligned dust grains are associated with clouds, we should use the data corresponding to the clouds. Zeeman measurements in Heiles (1989) provide $B \approx 1.2 \cdot 10^{-5}$ G for $n \approx 7$ cm$^{-3}$. This provides $t_r \approx 3 \cdot 10^{12}$ s, while $t_{gas} \approx 6 \cdot 10^{12}$ s, which gives for spherical grains $t_{gas}/t_r = 2$. If one takes into account that grains are non-spherical (Purcell & Spitzer 1971) and subject to additional disorientation through emission, $\delta_{eff} \sim 0.8$ can characterise alignment for $t_L < t_d$. This example of Heiles data for morphologically distinct shells may be considered to be extreme. However, numerous studies (see Myers 1985, 1987, Myers & Goodman 1988) indicate that molecular clouds are magnetically supported and exhibit motions which are supersonic 'in all but the smallest cloud regions' (Myers 1985). These motions are usually attributed to Alfvénic turbulence (see Aron & Max 1975, Falgarone & Puget 1986). However, the Alfvénic velocity is equal to the thermal velocity of atoms for 'standard' values of $B = 2 \cdot 10^{-6}$ G and $n = 20$ cm$^{-2}$. Therefore to reconcile widely accepted explanations of the molecular cloud support and non-thermal line widths with observations, one should increase substantially $B^2/n$ ratio. Note that it is this ratio that enters the expression for $t_{gas}/t_r$ and therefore an increase of Alfvénic velocity above the thermal one entails an increase of $\delta_{eff}$. If the Alfvénic velocity is five times greater than the thermal velocity $\delta_{eff}$ increases twenty five times as compared with 'standard' values. Even a less increase of $\delta_{eff}$ is adequate to explain the polarization if $t_L$ is several times greater than $t_{gas}$ as it is discussed in section 3.

Note, that the decrease of $B^2/n$ ratio as it follows from the data in Myers & Goodman (1988) may be one of the explanations for the decrease of polarization in dark clouds as indicated in Goodman (1994).

Therefore the cumulative effects of inhibited poisoning as compared to that assumed Spitzer & McGlynn (1979) and enhanced values of magnetic field from recent observations seems to favour the Purcell's mechanism to account at least for part of observational data. Indications that the degree of alignment correlates with the total magnetic field strength (Jones 1989), also favour such an interpretation as compared to the one where only grains with superparamagnetic inclusions are aligned and, if aligned, their alignment is perfect (Mathis 1986).

The 'standard' values of the ratio $B^2/n$ appeared at the time when the Zeeman effect measurements were giving only upper limits of the magnetic field strength. According to Heiles (1989), this was an unfortunate result of the assumption that the splitting is easier to discover for narrow lines. In fact, strong magnetic fields usually correspond to broad lines. Present-day data both on Zeeman splitting and non-thermal linewidths tend to suggest that the 'standard' value of $B^2/n$ probably underestimate the actual value of this ratio. Having the long-lived spin-up with $t_L$ within the range discussed in section 3, we feel quite comfortable with less than an order of magnitude increase of $B^2/n$ ratio as compared with its 'standard' value.

Note that we are concerned with the ratio $B^2/n$. For instance, using the 'standard' values of $B = 3 \cdot 10^{-6}$ G but $n = 5$ cm$^{-3}$, one obtains $t_{eff}/t_r \approx 0.1$, which is sufficient for explaining the polarization for the $t_L/t_{gas}$ ratio that we discussed.

To summarise, the Purcell's mechanism seems to be adequately strong in the view of recent observational and theoretical results. Its implications for grains of different sizes and chemical composition are discussed further on.

### 4.2  Small and large grains

Using the MNR grain model (Mathis et al. 1977), it is possible to show (Mathis 1979) that a good fit of the polarization curve may be obtained if grains with radii above a certain threshold are aligned. The above dependence was accounted for in Mathis (1986) by assuming that only those grains that contain at least one superferromagnetic cluster are being aligned. Within the adopted model, the probability of having such a cluster is greater for larger grains. Alternatively, for the Gold-type alignment (Lazarian 1994a), it is possible to show that only grains larger than $10^{-5}$ cm are sufficiently inertial to be aligned under Alfvénic perturbations in the typical ISM conditions.§ Here we are interested whether the Purcell's alignment can reproduce this type of behaviour for the polarization curve.

It was noted in section 2 that if the density of active sites over the grain surface is low, the absolute majority of small grains would not have sites of H$_2$ formation and therefore would not be aligned by the Purcell's mechanism. However, we do not believe that this is a probable explanation for the difference in the alignment of small and large grains.

Our calculations in section 3 have shown that poisoning intensifies for $l$ less than $l_{cr\ p} \sim 0.6 \cdot 10^{-5}$ cm. This critical size can further increase due to the inverse dependence of $T_s$ on $l$ (Greenberg 1971):

$$T_s(l) \approx T_{s0} \hat{l}^{-0.2} \qquad (17)$$

This is a result of suppressed thermal emission from small grains in infrared. Although the power-law index is just 0.2, this dependence may be essential in view of exponential dependence of $l_{cr\ p}$ on temperature [see Eq. (16)]. Thus, small grains are expected to be warm enough to have $l_{cr\ p} \sim 10^{-5}$ cm.

As soon as $l < l_{cr\ p}$, rapid poisoning of active sites inevitably causes $t_L \ll t_{gas}$. This forces down both the degree of suprathermality and the measure of alignment. One may erroneously conclude, that inspite of this drop, the degree of alignment has a formal limit of unity as $l$ tends to zero. However, $t_r \sim l^2 K(\Omega)$ and $K(\Omega) \sim \Omega$ [see Eq. (5)] in the limit of high velocities, $t_r$ may be found to be inversely proportional to $l$ if rotation is suprathermal and $l < l_{cr\ p}$. At the same time, if poisoning of active sites for small grains is inhibited, then starting from $l_{cr\ m}$ $t_r$ becomes independent of $l$. Another important factor that suppresses the alignment of small grains is their disorientation due to thermal emission. In fact, one can introduce a critical size¶

$$l_{cr\ t} \sim 10^{-6} \hat{n}_H^{-5/4} \text{ cm} \qquad (18)$$

starting from which the disorientation due to thermal emission dominates. Due to the dependence of $t_{rad}$ on $T_s^4$ and the change of grain temperature with $l$ according to Eq. (17), the resultant dependence of $t_x \approx t_{rad}$ is $l^{1.8}$. Therefore for small grains $l < l_{cr\ p}$ and $l < l_{cr\ m}$, $\delta_{eff} \sim l^{2.3}$.

At the same time, if rotation for small grains is not fast enough for their paramagnetic susceptibility to come to saturation, the degree of alignment increases with the decrease of size as $l^{-0.2}$ within the interval $l_{cr\ m} < l < l_{cr\ p}$. A transition from the long-lived spin-up to short-lived spin-up corresponds to a considerable drop in $\delta_{eff}$ and therefore the alignment of grains with $l < l_{cr\ p}$ is expected to be marginal unless very high values of the ratio $B^2/n$ are present in the grain environment. Both the disorientation through radiation and saturation of paramagnetic dissipation suppress the alignment for the smallest grains. Our computations show that we should expect the said drop for grains less than $10^{-5}$ cm.

---

§ Grains are expected to be aligned with their long axes perpendicular to magnetic field lines, which corresponds to observations (Lazarian 1994a).

¶ This is not an exact result as grain emissivity deviates substantially from the black body approximation for small grains. However, computations in Purcell & Spitzer [1971] show that for small grains disorientation due to thermal emission dominates.

|  | $l > l_{cr\ m}$ $l > l_{cr\ t}$ | $l < l_{cr\ m}$ $l > l_{cr\ t}$ | $l > l_{cr\ m}$ $l < l_{cr\ t}$ | $l < l_{cr\ m}$ $l < l_{cr\ t}$ |
|---|---|---|---|---|
| $l > l_{cr\ p}$ $l > l_{cr\ f}$ | $l^{-2/3}$ | $l^{4/3}$ | $l^{-2/3}$ | $l^{4/3}$ |
| $l > l_{cr\ p}$ $l < l_{cr\ f}$ | $l^{-2}$ | $l^0$ | $l^{-2}$ | $l^0$ |
| $l < l_{cr\ p}$ $l > l_{cr\ f}$ | $l^{1/3}$ | $l^{17/6}$ | $l^{17/15}$ | $l^{109/30}$ |
| $l < l_{cr\ p}$ $l < l_{cr\ f}$ | $l^{-1}$ | $l^{3/2}$ | $l^{-1/5}$ | $l^{23/10}$ |

**Table 1.** The dependence of $\delta_{eff}$ on grain size for various combinations of the critical sizes if the number of active sites is proportional to the grain surface. $l_{cr\ m}$ = critical size below which the paramagnetic susceptibility is saturated [Eq. (6)], $l_{cr\ f}$ = critical size above which a partial preservation of alignment in the course of crossovers is important [Eq. (9)], $l_{cr\ p}$ = critical size below which the oxygen poisoning of active sites is enhanced [Eq. (16)] $l_{cr\ t}$ = critical size below which disorientation due to thermal emission dominates [see Eq. (18)].

|  | $l > l_{cr\ m}$ $l > l_{cr\ t}$ | $l < l_{cr\ m}$ $l > l_{cr\ t}$ | $l > l_{cr\ m}$ $l < l_{cr\ t}$ | $l < l_{cr\ m}$ $l < l_{cr\ t}$ |
|---|---|---|---|---|
| $l > l_{cr\ p}$ $l > l_{cr\ f}$ | $l^0$ | $l^2$ | $l^0$ | $l^2$ |
| $l < l_{cr\ p}$ $l < l_{cr\ f}$ | $l^1$ | $l^{21/6}$ | $l^{27/15}$ | $l^{129/30}$ |

**Table 2.** The dependence of $\delta_{eff}$ on grain size when the number of active sites is stabilised at the level $\nu_{cr}$.

For large grains, partial preservation of orientation during crossovers becomes important and this improves their alignment: for $l > l_{cr\ m}$ and $l < l_{cr\ m}$, $\delta_{eff}$ becomes proportional to $l^{-2/3}$ and $l^{4/3}$, respectively. Other situations that may emerge for various combinations of four critical sizes and $\nu < \nu_{cr}$ are shown in Table 1, whereas the dependencies for $\nu > \nu_{cr}$ are given in Table 2. This information may be used for future quantitative comparison of theory and observations.

Although possible in some situations, not all the combinations presented in Tables 1 and 2 are equally probable for diffuse clouds. We believe that it is more likely that there $\delta_{eff}$ starts from $\sim 0$ for the smallest grains and so as $l^{23/10}$ until $l = l_{cr\ m}$, after which it changes to $l^{-1/5}$ A considerable increase of $\delta_{eff}$ follows at $l = l_{cr\ p}$, which corresponds to the transition to a long-lived spin-up. A decrease of $\delta_{eff}$, which is proportional to $l^{-2}$, at $l = l_{cr\ f}$ becomes less rapid ($\sim l^{-2/3}$) due to a partial preservation of alignment during crossovers. When the number of active sites exceeds $\nu_{cr}$ and the decrease of $\delta_{eff}$ stops.

It may be considered strange, but in view of new observational results concerning magnetic fields in diffuse clouds Heiles (1989), Myers & Goodman (1988), Heiles *et al.* (1993) it may become more important to explain marginal alignment of small grains rather than to account for good alignment of large grains. Indeed, both the Purcell's alignment as it is discussed here and superparamagnetic alignment as it was suggested in Mathis (1986) fail to explain marginal alignment of smaller grains if the ratio $B^2/n$ is large enough.

To summarise, the Purcell's mechanism is able to reproduce qualitatively the observed features of the polarization curve. A more quantitative study will be done elsewhere.

### 4.3 Grains of different chemical composition

There exist a considerable convergence in views on the dielectric nature of polarising particles (see e.g. Whittet 1993), although the values of the imaginary part of the grain index of refraction is a subject of controversy. Our results indicate, that 'H$_2$ driven' suprathermal rotation and therefore the Purcell's alignment are efficient for grains with surfaces of silicate, water ice or polymeric hydrocarbon. As the IR polarimetry shows substantial polarization in both silicate 9.7 $\mu$m feature and 3.7 $\mu$m water ice band (e.g. Whittet 1992) while graphite and amorphous carbon grains are likely not to be aligned (Mathis 1986), this may serve as an argument in favour of the Purcell's mechanism. Grains with surfaces of aromatic carbonaceous material and graphite should rotate thermally and therefore be subjected to the Davis-Greenstein alignment. But with the known ratios $B^2/n$ this alignment is marginal. If it is suprathermal rotation that controls the structure of grains (Wright 1994), the above grains are expected to be fluffy and therefore will not be aligned by Alfvén-Gold processes discussed in Lazarian (1994a).

## 5 TEST CASES
### 5.1 Cold and warm grains

We have seen that due to temperature increase, poisoning atoms become sufficiently mobile to attack active sites directly and therefore the life time of active sites decreases. Our results show that for icy grains an increase of temperature up to 20 K leads to such a considerable increase in poisoning that long-lived spin-up becomes impossible. If the temperature increases further, H$_2$ formation becomes suppressed. Indeed, for a reaction between physically and chemically adsorbed H atoms to occur, an activation barrier should be overcome. A critical temperature $T_{cr}$ can be defined for which the probability of reaction with a chemisorbed atoms is equal to the rate of evaporation:

$$T_{cr} = \frac{E_h}{k \ln(\tau_r \nu_0)} \qquad (19)$$

where $\tau_r$ is the reaction time-scale $\sim 10^{-6}$ s. Estimates provide $T_{cr} \approx 50$ K for $H_2$ formation over a silicate surface (Tielens & Allamandola 1987). Above the critical temperature, the reaction rate drops exponentially and the reduction factor at $T_s \sim 75$ K is expected to be $\sim 5 \cdot 10^{-3}$. Such a decrease in efficiency of $H_2$ formation entails the same decrease in attainable angular velocity $\Omega$ making it comparable with the velocity of thermal rotation $\omega_T$. Therefore the Purcell's mechanism becomes inapplicable. Our discussion in section 3 shows that even smaller increase of temperature makes long-lived spin-up impossible for grains with mantles of amorphous ice. This means that, for instance, the aligned warm ($T_s \sim 75$ K) dust in the core regions of OMC-1 (Chrysostomou et al. 1994) can not rotate suprathermally due to $H_2$ formation. In general, it is unlikely that the suprathermal paramagnetic alignment of warm ($T_s > 50$ K) dust may be efficient. It is either the Gold-type alignment (see Lazarian 1994a) or Davis-Greenstein mechanism that governs the alignment of warm dust. To distinguish between the two cases one needs to study the polarization curve. The Davis-Greenstein mechanism should result in efficient alignment of small grains,$^\|$ while Gold-type processes preferably align larger grains as a result of Alfvénic phenomena.

### 5.2 H-atom population

It is easy to see from Eqs. (1)–(3) that $\Omega$ is proportional to $n_H/n$ ratio and therefore becomes of the order of thermal angular velocity as soon as $n_H/n$ ratio becomes as small as $10^{-3}$. Such circumstances are expected in dark clouds, where the UV is totally excluded and the residual H-atom population is maintained by cosmic rays; the steady state value of $n_H$ is $\sim 1$ cm$^{-3}$ (Duley & Williams 1984). However, there are indications (Norman & Silk 1980, Goldsmith et al. 1986) that this steady state may not be reached in many of dark clouds due to gas cycling between low and high density phases. For instance, $n_H$ in the range of 50-160 cm$^{-3}$ is reported in Snell (1981) for several typical dark clouds which provides an estimate for $n_H/n$ in the range of $2 \cdot 10^{-2} - 5 \cdot 10^{-4}$, which can be interpreted as the evidence in favour of suprathermal rotation of grains in some of dark clouds. In general, suprathermal rotation and therefore the Purcell's alignment is expected to be more efficient in outer part of dark clouds, where $n_H/n$ is greater. This qualitatively corresponds to the results in Goodman (1994). In fact, we believe that both the decrease of $B^2/n$ ratio as a result of ambipolar diffusion as well as $n_H/n$ ratio contribute to the decrease of alignment in dark clouds.

### 6 CONCLUSIONS

A range of applicability of the Purcell's mechanism has been defined in view of new results in the field of grain physics and chemistry. It has been shown that grains with amorphous $H_2O$ ice mantles, defected silicate and polymeric carbonaceous surfaces are likely to rotate suprathermally and can be aligned by the said mechanism if the relative abundance of atomic hydrogen $n_H/n$ is not less than $10^{-3}$ and grains are cool ($T_s < 50$ K). At the same time, grains covered with aromatic carbonaceous material or graphite are not expected to rotate suprathermally due to $H_2$ formation and therefore this mechanism is not applicable to them. It has been shown that the Purcell's mechanism provides a preferential alignment of large grains, while only a marginal alignment of grains with $l < 10^{-5}$ cm is expected.

---

$^\|$ This is true unless the grains are Mathis' grains (1986), which population of larger grains is characterised by greater magnetic susceptibility. We will discuss this interesting proposal in our next paper.

### Acknowledgements

It is a great pleasure to express here my gratitude to J. Mathis, communications with whom stimulated my study of the difference in alignment of large and small grains. His comments also contributed to substantial improvement of the original version of the paper. This research also owes much to my discussions with P. Myers and D. Williams. Encouragement of N. Weiss and communications with J. Greenberg & E. Wright are gratefully acknowledged. I would also like to thank the Institute of Astronomy, University of Cambridge, for the Isaac Newton Scholarship they generously awarded me, which provided the financial support for this work.

## APPENDIX A1: DISORIENTATION PARAMETER

A partial preservation of alignment during crossovers can be described using parameter $F$ introduced in Spitzer & McGlynn (1979) as $\cos\langle\chi_T\rangle = \exp(-F)$, where $\chi_T$ is the angle of deviation of the angular. Calculations in Spitzer & McGlynn (1979) give

$$F = \frac{\pi}{4}\left\{\frac{\langle(\triangle J_z)^2\rangle}{|J_\perp|\langle\triangle J_z\rangle}\left(1 + \frac{3\langle(\triangle J_\perp)^2\rangle}{2\langle(\triangle J_z)^2\rangle}\right)\right\} \quad (A1)$$

in the approximation of the constant angular momentum component that is perpendicular to the axis of the greatest inertia. Note, that $\triangle J_z$ and $\triangle J_\perp$ are elementary changes of the angular momentum in an individual torque event, and $J_\perp$ is the value of the component of angular momentum perpendicular to the axis of major inertia.

The squared dispersion of $J_\perp$ at $t = 0$ is given in Spitzer & McGlynn (1979, eq. 42):

$$\langle J_\perp^2(0)\rangle \approx \left(\frac{3}{2}\right)^{1/3}\Gamma\left(\frac{4}{3}\right)(A_aN)^{1/3}\langle\triangle J_z^{-2/3}\rangle I_z^{2/3}$$
$$\times \langle(\triangle J_\perp)^2\rangle \quad (A2)$$

where $\Gamma(x)$ is the Gamma function, and $A_a$ is the coefficient of the Barnett relaxation (Purcell 1979, eq. 46) and $N \approx \gamma l^2 n_H v_H$ is the frequency of the torque events.

To calculate $F$ from Eq. (A1), one needs to know the average of $J_\perp^{-1}(0)$. Note, that $\langle J_\perp^2(0)\rangle$ is a sum of $\langle J_x^2(0)\rangle$ and $\langle J_y^2(0)\rangle$ and due to the symmetry inherent to the problem

$$\sigma_1^2 = \langle J_x^2(0)\rangle = \langle J_y^2(0)\rangle = 0.5\langle J_\perp^2(0)\rangle \quad (A3)$$

For a Gaussian distribution, one gets

$$\frac{1}{\langle J_\perp(0)\rangle} = \frac{1}{2\pi\sigma_1^2}\iint_{-\infty}^{+\infty}\frac{1}{\sqrt{x^2+y^2}}$$
$$\times \exp\left\{-\frac{x^2+y^2}{2\sigma_1^2}\right\}\,dx\,dy \quad (A4)$$

which can be calculated in polar coordinates to give

$$\frac{1}{\langle J_\perp(0)\rangle} = \sqrt{\frac{\pi}{2}}\frac{1}{\sigma_1} = \frac{\sqrt{\pi}}{\langle J_\perp^2(0)\rangle^{1/2}} \quad (A5)$$

Similarly the Gaussian averages can be found for $\langle\triangle J_z^{-2/3}\rangle$. The dispersion of $\langle\triangle J_z\rangle$ which we will denote $\sigma_2$ is inversely proportional to the square root of the number of active sites $\nu$ over the grain surface and depends on grain geometry and distribution of active sites in respect to the $z$-axis. For a disc shaped grain $\sigma_2 = 0.25 l m_{H_2} v_{H_2} \nu^{-1/2}$. Therefore

$$\langle\triangle J_z^{-2/3}\rangle = \frac{1}{\sigma_2\sqrt{2\pi}}\int_{-\infty}^{+\infty}\frac{1}{x^{2/3}}\exp\left\{-\frac{1}{2}\frac{x^2}{\sigma_2^2}\right\}\,dx \quad (A6)$$

which gives (see Gradstein & Ryzhik, [3.462(1)] and [9.241(1)])

$$\langle\triangle J_z^{-2/3}\rangle = \frac{1}{\sigma_2^{\frac{2}{3}}}\left\{\sqrt{\frac{2}{\pi}}\Gamma\left(\frac{1}{3}\right)D_{-1/3}(0)\right\} \quad (A7)$$

where $D_p(x)$ is the parabolic cylinder function and $\Gamma(x)$ is the Gamma function.

Thus substituting Eq. (A7) into Eq. (A2)

$$F = \tilde{\kappa}\frac{\langle(\triangle J_z)^2\rangle}{(A_aN)^{1/6}I_z^{1/3}\langle(\triangle J_z)^2\rangle^{1/3}\langle(\triangle J_\perp)^2\rangle^{1/2}}$$

$$\times \quad \left(1 + \frac{3\langle(\triangle J_\perp)^2\rangle}{2\langle(\triangle J_z)^2\rangle}\right) \tag{A8}$$

where the numerical factor is equal to

$$\tilde{\kappa} = \left(\frac{3}{4}\right)^{-1/6} \Gamma^{-1/2}(4/3) \int_0^\infty e^{-\frac{x^2}{2}} x^{-2/3} dx \tag{A9}$$

Assuming $\langle(\triangle J_z)^2\rangle = \langle(\triangle J_\perp)^2\rangle$, one obtains

$$F = \frac{5}{2}\tilde{\kappa} \frac{\langle(\triangle J_z)^2\rangle^{1/2}}{(A_a N)^{1/6} I_z^{1/3} \langle[\triangle J_z]^2\rangle^{1/3}} \tag{A10}$$

Substituting $\langle(\triangle J_z)^2\rangle = \frac{l^2}{9} m_{H2}^2 v_{H2}^2$ as well as expressions for $A_a$, $N$ and $\langle[\triangle J_z]^2\rangle$ into Eq. (A10), one gets

$$F \approx 0.8 \; (\hat{v}_{H2}\hat{\nu})^{1/3} \hat{\varrho}^{-1/2} \hat{l}^{-2} (\hat{n}_H \hat{v}_H \hat{\gamma}_1)^{-1/6} \tag{A11}$$

where $\nu$ is normalised by 100. If the number of active sites $\nu$ over the grain surface is assumed to be proportional to the surface area of the grain $S_g(l)$, $F$ decreases as $l^{-4/3}$ with the increase of grain size. If the number of active sites is stabilised due to active poisoning at the level $\nu_{cr}$, $F$ decreases as $l^{-2}$.